\documentclass[floatfix,twocolumn,showpacs,nofootinbib,prd,aps,tightenlines]{revtex4-1}
\usepackage{graphicx}
\usepackage{mathrsfs}
\usepackage{psfrag}
\usepackage{dcolumn}
\usepackage{bm}
\usepackage{amsmath}
\usepackage{bm}
\usepackage{appendix}
\usepackage[caption=false]{subfig}

\newcommand{\scri}{\mathscr{I}}

\newcommand{\bea}{\begin{eqnarray}}
\newcommand{\eea}{\end{eqnarray}}
\newcommand{\beq}{\begin{equation}}
\newcommand{\eeq}{\end{equation}}

\usepackage{color}

\begin{document}

\title{The maximum radiated energy and final spin of high speed collision of two black holes}

\author{James Healy}
\author{Alessandro Ciarfella}
\author{Carlos O. Lousto}
\affiliation{Center for Computational Relativity and Gravitation,
School of Mathematical Sciences,
Rochester Institute of Technology, 85 Lomb Memorial Drive, Rochester,
New York 14623, USA}

\date{\today}

\begin{abstract}
We performed a series of 769 full numerical simulations of high
energy collision of black holes to search for the maximum gravitational energy
emitted $E_{rad}$, during their merger.  We consider equal mass binaries with
spins pointing along their orbital angular momentum $\vec{L}$ and perform a search
over impact parameters $b$ and initial linear momenta $p/m=\gamma v$ to
find the maximum $E_{rad}$ for a given spin $\vec{S}$. The total
radiated energy proves to have a weak dependence on the intrinsic spin $s$ of the holes, 
for the sequence $s=+0.8, 0.0, -0.8$ studied here. We thus estimate the maximum
$E_{rad}^{max}/M_{ADM}\approx32\%\pm2\%$ for these direct merger encounters.
We also explore the radiated angular momentum and the maximum spin
of the merger remnant (within these configurations), finding $\alpha_f^{max}=0.987$.
We then use the zero frequency limit expansion to analytically model the radiated
energy in the small impact parameter and large initial linear momentum regime.
\end{abstract}

\pacs{04.25.dg, 04.25.Nx, 04.30.Db, 04.70.Bw}\maketitle

\section{Introduction}\label{sec:Intro}

The extreme scenario of high energy collision of
black holes has been the subject of detailed studies
in the realm of high-energy colliders to help discover the
fundamental laws of nature \cite{Cardoso:2012qm,Berti:2016rij}, with
applications to the gauge/gravity duality and holography
\cite{Cardoso:2014uka}, as well as
tests of the radiation bounds theorems and cosmic censorship
conjecture in General Relativity
\cite{Hawking:1971tu,Eardley:2002re,Siino:2009vw}, and with
regards to primordial black hole collisions in the early
universe \cite{Franciolini:2022htd,Ding:2020ykt,Cai:2019igo}.

The high energy scenario was studied with full numerical techniques
in \cite{Sperhake:2008ga} to compute the maximum energy
radiated by equal mass, nonspinning black holes in an
ultrarelativistic headon collision. This study was then followed
up \cite{Sperhake:2012me} by the claim that the spin effects did
not matter for these collisions.  In \cite{Healy:2015mla} we
revisited the headon scenario using our new initial data
\cite{Ruchlin:2014zva} with low spurious initial radiation content
for more accurate estimates of the maximum energy radiated,
placing it at about $13\%$.  Non headon high energy collisions have
been studied in \cite{Shibata:2008rq} and then with notable analytic detail
in \cite{Berti:2010ce}.  Some of the early reviews on the subject are
\cite{Cardoso:2012qm,Berti:2016rij} and a more up-to-date one in
\cite{Sperhake:2019oaw,Bozzola:2022uqu,Page:2022bem}. Also, 
interestingly, the scattering of two nonspinning black holes
may lead to their spinup during their short term interactions
\cite{Nelson:2019czq}.

In a previous work \cite{Healy:2022rng} we explored the maximum recoil
velocity of such high energy black hole encounters by performing a
search in the 4-dimensional parameter space of impact parameters $b$,
linear momentum (per horizon mass) $\gamma v$, and spins magnitude
and orientations in the orbital plane, finding the maximum recoil velocity
of the final merged black hole to be about $10\%$ the speed of light.

Here we come back to this high energy collision scenario and target our
studies to the search for the maximum achievable radiated gravitational
energy from these grazing, high energy collisions and merger of binary black holes,
where now the spin orientation of the holes play a different
(as we will see, less relevant) role than in the computation of the recoil. 
We will perform supercomputer simulations
by directly solving fully numerically the General Relativity
field equations and supplement their results with analytic computations
in the zero frequency limit (ZFL) \cite{Smarr:1977fy,Berti:2010ce} 
to fit those results.

\section{Techniques}\label{sec:Tech}

In order to perform the full numerical simulations we have used the
LazEv code\cite{Zlochower:2005bj} with 8th order spatial finite
differences \cite{Lousto:2007rj}, 4th order Runge-Kutta time
integration with a Courant factor $(\Delta t/\Delta x=1/4)$
implementation of the moving puncture
approach \cite{Campanelli:2005dd}.  To compute the numerical
(Bowen-York) initial data, we use the {\sc TwoPunctures}
\cite{Ansorg:2004ds} code.  We use the {\sc AHFinderDirect} code
\cite{Thornburg2003:AH-finding} to locate apparent horizons and
measure individual masses $m_H$ and the magnitude of the horizon
spin $S_H$, using the {\it isolated horizon} algorithm as implemented in
Ref.~\cite{Campanelli:2006fy}.

The {\sc Carpet} mesh refinement driver provides a ``moving boxes''
style of mesh refinement.  The grid structure of our mesh refinements
have a size of the largest box for all simulations of $\pm400M$.  The
number of points between 0 and 400 on the coarsest grid is 100 for a
resolution of n100.  So, the grid spacing on the coarsest is
$4M$.  The resolution in the wavezone (two refinement levels below
the coarsest) is $1M$ (i.e. n100 has $M/1.00$).
The grid around the (equal mass) black holes is fixed at
$\pm0.3M$ in size and is the 9th refinement level.  Therefore the grid
spacing at the finest resolution is $M/256$.

The extraction of gravitational radiation from the numerical
relativity simulations is performed using the formulas (22),(23),(24)
from \cite{Campanelli:1998jv} for the energy and linear and angular
momentum radiated, respectively, in terms of the extracted Weyl scalar
$\Psi_4$ at the observer location sphere $R_{obs}=113M$.  We then analytically
\cite{Nakano:2015pta} extrapolate expressions for $r \psi_4$ from this
finite observer location $R_{obs}$ to $\scri^+$.  To compute Energy,
angular and linear momenta radiated from the waveforms we will subtract, in
the time domain, at a post-processing stage, the initial burst of spurious Bowen-York
data radiation content reaching the observer, to thus obtain more physical
values.

\section{Numerical Results}\label{sec:Results}

In \cite{Healy:2022rng} we have explored the maximum the recoil from
equal mass binary black holes with opposite spins on the orbital
plane. We have designed simulations to explicitly model the problem in
terms of the Bowen-York initial linear momentum per horizon mass
of the holes, $\gamma v=p/m_{H}$, (with $\gamma=(1-v^2)^{-1/2}$, the
Lorentz factor, and the horizon area $A_H=16\pi m_{irr}^2$ providing
$m_H^2=m_{irr}^2+{S}_H/4m_{irr}^2$), dimensionless (per unit mass $M$)
impact parameter, $b$, and spin, $\vec{s}=\vec{S}_H/m_H^2$ (where
$m_H=m_{1,2}$ is the horizon mass of each hole) magnitude and
orientation, i.e. a four dimensional parameter search.

\begin{figure}
\includegraphics[angle=0,width=\columnwidth]{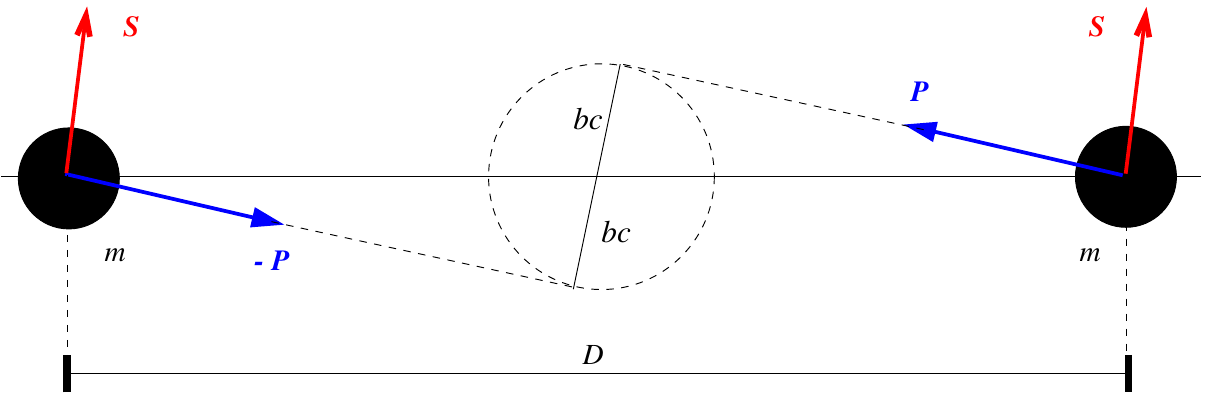}
\caption{Initial configurations of the high energy collisions of black holes.
 With $s=s^z=0.0,\pm0.8$ and $D=50M$.
  \label{fig:IDE}}
\end{figure}

Here we will supplement that study of black holes high energy collisions
by a search of the maximum
radiated energy (and final black hole spin).
Our simulations families consist of equal mass black holes
with spins aligned (or counter-aligned) with the orbital angular momentum
$\vec{L}$, and of an initial spin magnitude, $s=0.0, \pm0.8$, for each
of the holes as displayed in Fig.~\ref{fig:IDE}.
We will then explore different initial relativistic velocities
(or momentum per mass), $\gamma v$, and
(normalized by $M$) impact parameters, $b$,
as measured from the initial separation of the holes, $D=50M$
(with $M=M_{ADM}$, the total Arnowitt-Deser-Misner (ADM) mass of
the system~\cite{Arnowitt:1962hi}).
Note that here $m_1+m_2=M_{ADM}-E_b$, the addition of the horizon
masses of the system, does not add up to 1 (as was normalized in
\cite{Healy:2022rng}), but here differs from the normalized
$M_{ADM}=1$ by the (negative) initial binding energy, $E_b$.

The value for the impact parameter $b_{max}$ that leads to the
maximum energy radiated (or maximum spin of the merged black hole)
corresponds closely to the critical value of the impact parameter
$b_{c}$ separating the direct merger from the scattering of the holes.
Those two families being separated by a set of elliptic orbits.
A similar analysis can be done to complete this two parameters
search by varying the initial velocity (or
linear momentum per horizon mass) of the holes, $\gamma v=p/m_{H}$,
maximizing energy and final black hole spin for values about the
critical initial linear momentum separating the direct merger from
scattering of the holes.

Typically, for our simulations,
we use a grid with global resolution n100 and
with 10 levels of refinement, the coarsest of which has
resolution of $4M$ and outer boundary of $r=400M$, with each successive
grid with twice the resolution. If 
we label the coarsest grid $n=0$, and the finest grid $n=9$, the resolution
on a given level is $M/2^{(n-1)}$.  The wavezone is $n=2$ with a resolution of
$M/1$ and boundary out to $r=125M$.  The finest grid has a resolution of
$M/256$ with a cube of sides of 1.0 centered around each black hole. 
For higher initial ${\gamma}v$, an additional refinement level is added
due to the decreasing black hole radius, with a resolution of $M/512$.
In \cite{Healy:2022rng} we reported several convergence studies of these
kind of simulations confirming the accuracy of our base grid and resolutions.

\begin{figure}
\includegraphics[angle=0,width=\columnwidth]{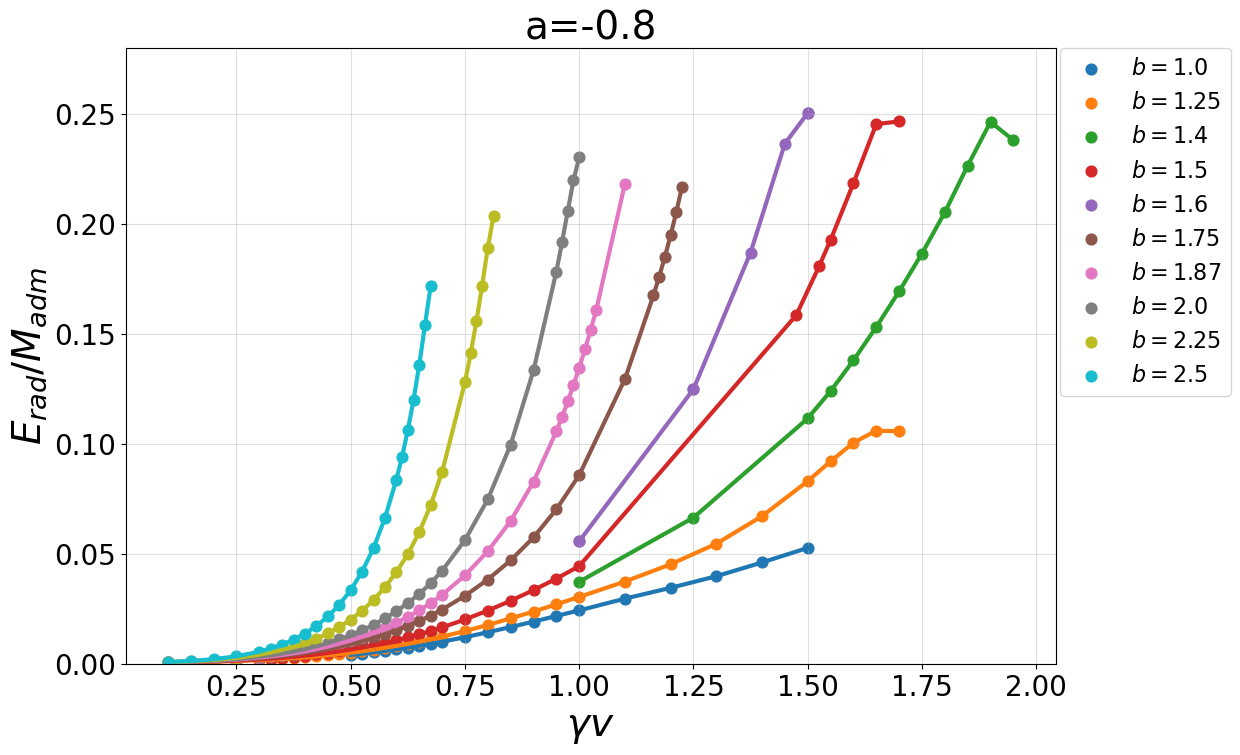}\\
\includegraphics[angle=0,width=\columnwidth]{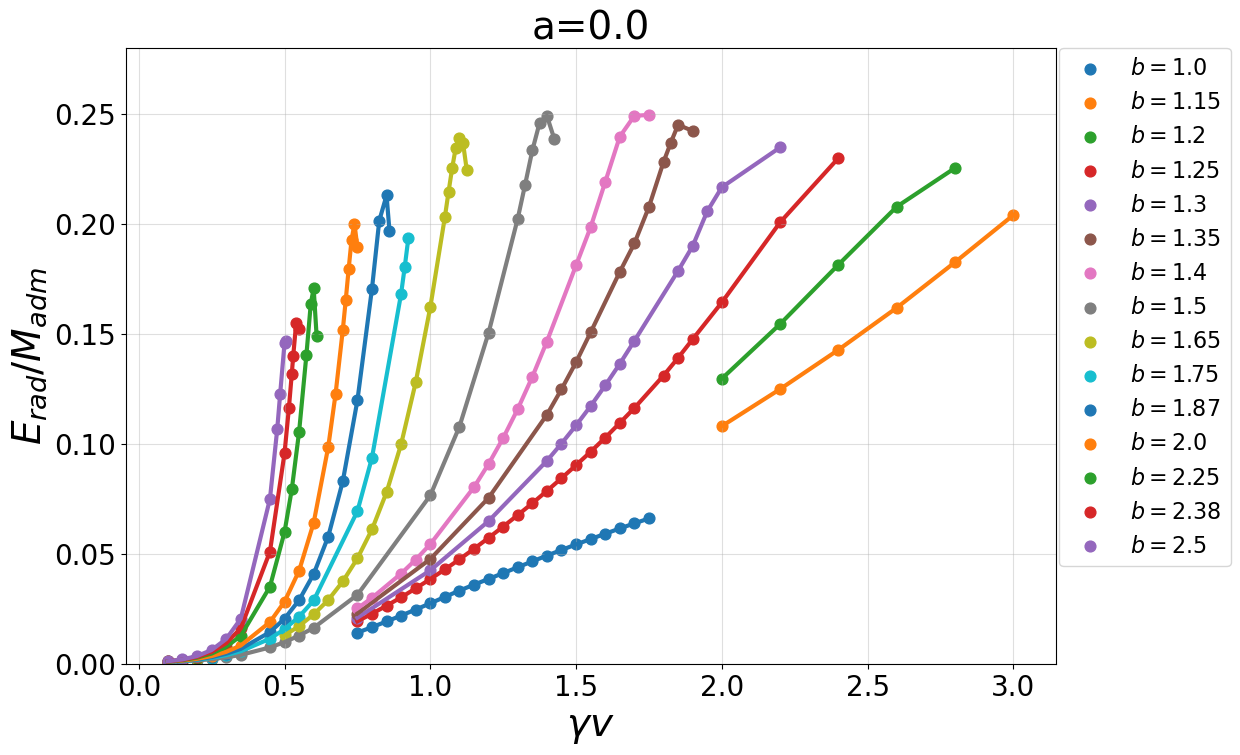}\\
\includegraphics[angle=0,width=\columnwidth]{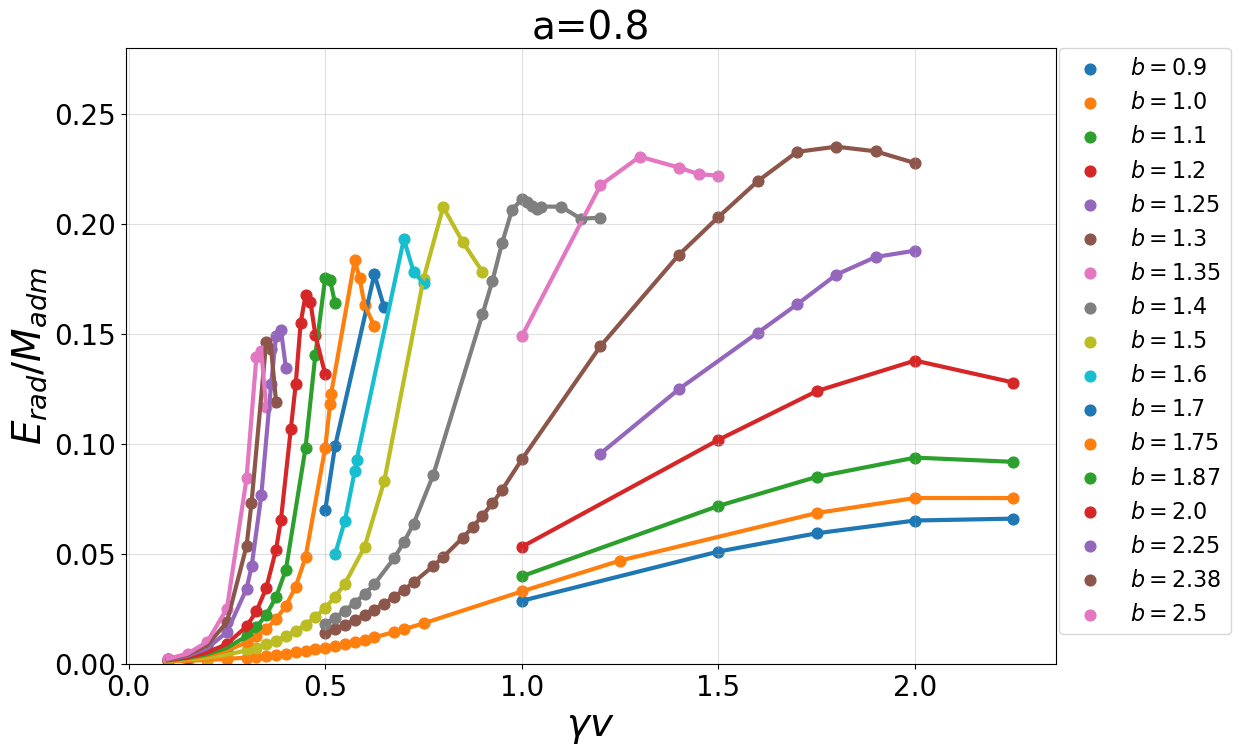}\\
  \caption{Maximum $E_{rad}/M_{ADM}$ for different initial momenta parameters $\gamma v$ and impact parameters $b$ for $s=-0.8, 0.0, +0.8$. 
  \label{fig:Evsgv}}
\end{figure}

Fig.~\ref{fig:Evsgv} displays the maximum radiated energy versus the initial
$\gamma v$ of the simulations and the impact parameters $b$ as curve levels
for $s=-0.8, 0.0, +0.8$. We have performed this search using 294, 253 and
222 simulations for spins $s=-0.8, 0.0, +0.8$ respectively.
We observe a relatively weak dependence on the values of the spins, particularly
for the counteraligned and zero spins while a lower maximum is found for
the aligned spins with respect to the orbital angular momentum.

\begin{figure}
  \includegraphics[angle=0,width=0.75\columnwidth]{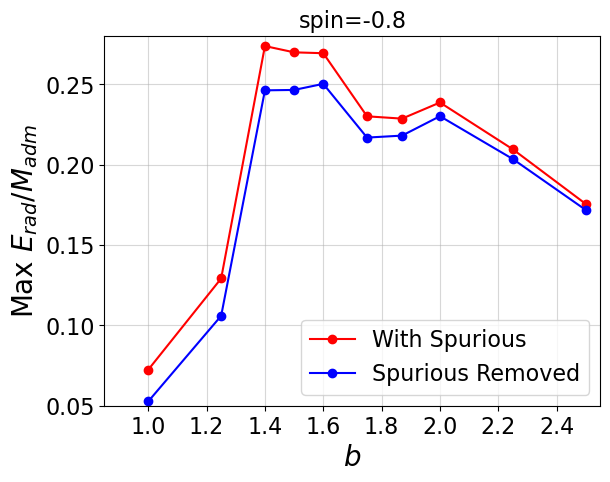}\\
  \includegraphics[angle=0,width=0.75\columnwidth]{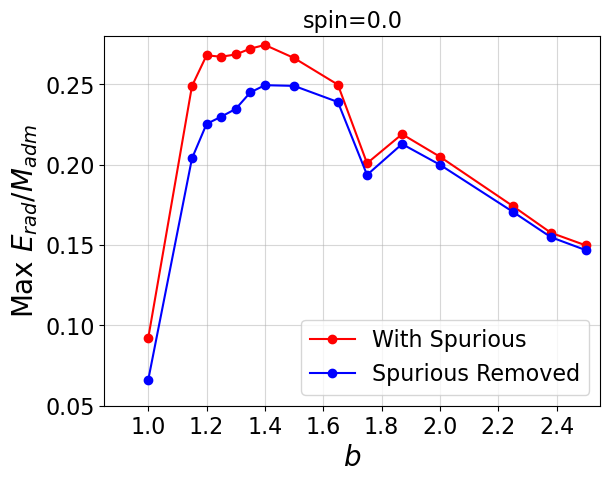}\\
  \includegraphics[angle=0,width=0.75\columnwidth]{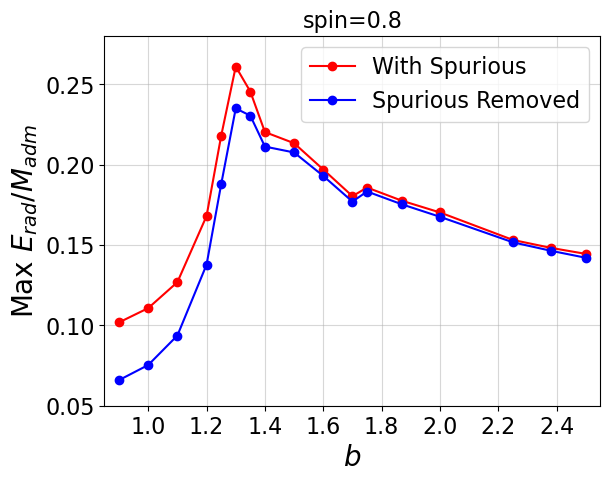}
  \caption{Maximum $E_{rad}/M_{ADM}$ with removal of the initial burst of radiation for different initial momenta parameters $\gamma v$ and impact parameters $b$ for $s=-0.8, 0.0, +0.8$. 
  \label{fig:EvsgvR}}
\end{figure}

In order to consider the effects of the initial data radiation content
in Fig.~\ref{fig:EvsgvR} we display the effects of the removal of
the initial burst of spurious radiation on the evaluation of
the maximum radiated energy. 

Table~\ref{tab:Emax-spurious} provides some numerical detail of the subtraction
of the spurious radiation from the initial data to extract the more physical
gravitational radiation from the high energy collision of the two black holes.

\begin{table}
  \caption{Corrections to the maximum $E_{rad}$ due to the initial spurious burst of radiation
\label{tab:Emax-spurious}}
\begin{ruledtabular}
\begin{tabular}{lllll}
spin & $b$ & $E_{rad}$ & $E_{rad}$-spurious & difference\\
\hline
-0.8  &   1.40 & \bf{0.273819}  & 0.246259  &-0.027560\\
-0.8  &   1.50 & 0.269910       & 0.246511       &-0.023399\\
-0.8  &   1.60 & 0.269365       & {\bf 0.250295} &-0.019070\\
-0.8  &   1.75 & 0.230099       &      0.216861  &-0.013238\\
\hline
0.0 & 1.30 & 0.268583 & 0.234700 &-0.033883\\
0.0 & 1.35 & 0.272209 & 0.244980 &-0.027229\\
0.0 & 1.40 & {\bf 0.274504} & {\bf 0.249464} &-0.025040\\
0.0 & 1.50 & 0.266362 & 0.249038 &-0.017324\\
0.0 & 1.65 & 0.249796 & 0.239010 &-0.010786\\

\hline
+0.8 & 1.25 & 0.217894 & 0.187761 &-0.030133\\
+0.8 & 1.30 & {\bf 0.261063} & {\bf 0.235021} &-0.026042\\
+0.8 & 1.40 & 0.220289 & 0.211261 &-0.009028\\

\end{tabular}
\end{ruledtabular}
\end{table}

We observe that while the initial burst of radiation is significant,
due to both, the presence of spin and relativistic velocities of the
holes being simply modeled by the conformally flat Bowen-York ansatz,
it is an order of magnitude smaller than the total energy radiated and
does not significantly shifts the values of $\gamma v$ and $b$ that lead
to its maximum.

\subsection{Numerical convergence targeted studies}\label{sec:convergence}

Given the large amount of simulations needed to establish the region of
parameters that may lead to the largest energy radiated we have assumed
a hierarchical approach by first locating the near region of the maximum
$E_{rad}$ in the $(b,\gamma v)$ parameter space with low resolution runs,
i.e.. n100, and then explore this region with targeted higher resolution
simulations that allow us to verify our hierarchical hypothesis as well as
providing us with a means to extrapolate in the convergence regime
to infinite resolution values.

\begin{figure}
  \includegraphics[angle=0,width=\columnwidth]{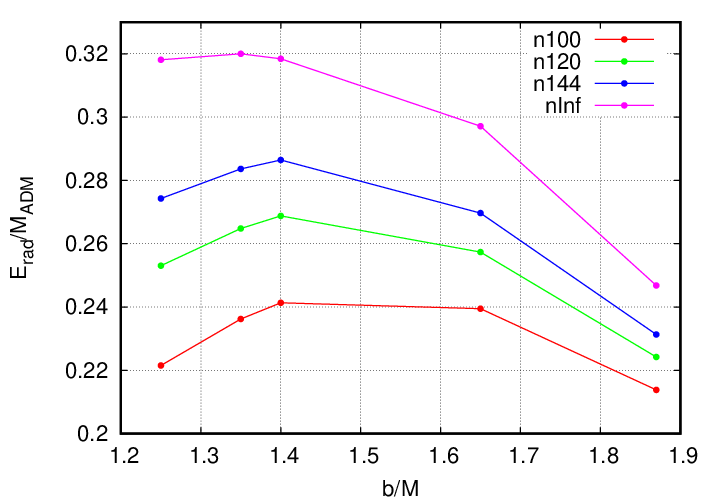}
  \caption{Maximum $E_{rad}/M_{ADM}$ vs. impact parameter $b/M$ for different numerical resolutions and its extrapolation to infinite resolution (ninf=$n_\infty$) for the case $s=0$.
  \label{fig:s0conv}}
\end{figure}

To explicitly display this procedure we consider the $s=0$ case by
supplementing the n100 simulations with n120 and n144 on by successively
increasing the global finite differences resolutions by factors of 1.2 .
We have adopted here an improved grid with one additional refinement level,
i.e. eleven, and changed the shift parameter $\eta$ from 2 to 1, hence
we rerun the n100 simulations together with the n120 and n144 to obtain
the convergence rates using formulas (5) in \cite{Lousto:2019lyf}.
This sequence is repeated for different values of the impact parameter
$b/M$ as displayed in Fig.~\ref{fig:s0conv}.

Once established the value of $b/M=1.35$ the analysis is
followed up by a verification of the $\gamma v$ dependence around
the $E_{rad}$ peak with results as displayed in table~\ref{tab:b135}.

\begin{table}
  \caption{Dependence of the maximum radiated energy on the parameter
    $(\gamma v)$ around $s=0$ maximum for different numerical resolutions
    (n100,n120,n144) and its extrapolation to infinite resolution n$_\infty$
    with a found convergence order.
    \label{tab:b135}}
\begin{ruledtabular}
\begin{tabular}{lcccccc}
$b/M$ & $(\gamma v)$  & n100 & n120 & n144 & n$_\infty$ & order\\
\hline
1.35 & 1.850 & 0.23254 & 0.26351 & 0.28297 & 0.3159 & 2.54\\
1.35 & 1.900 & 0.23620 & 0.26479 & 0.28363 & 0.3200 & 2.28\\
1.35 & 1.925 & 0.23635 & 0.26417 & 0.28272 & 0.3198 & 2.22\\
1.35 & 1.950 & 0.23206 & 0.25984 & 0.27851 & 0.3168 & 2.17\\
\end{tabular}
\end{ruledtabular}
\end{table}

A similar process described here for the $s=0.0$ case is also performed for the
$s=\pm0.8$ cases and the final results are summarized in table~\ref{tab:Emax}.
Thus totaling 30 additional runs in this targeted study in addition to the
original 739 of the upper hierarchical search.

\begin{table}
\caption{
  The maximum energy radiated $E_{rad}^{max}$ for spins $0.0,\pm0.8$.
  All simulations have equal mass $q=m_1/m_2=1$, and were initially placed at $x_{1,2}/M=\pm25$. 
  \label{tab:Emax}}
\begin{ruledtabular}
\begin{tabular}{rccccccc}
spin &
$b/M$ &
$(\gamma v)$  &
n100 &
n120 &
n144 &
n$_\infty$ &
d\\
\hline
-0.8 &
1.40 &
1.900 &
0.24000 &
0.26712 &
0.28587 &
0.3279 &
2.02\\
0.0 &
1.35 &
1.900 &
0.23620 &
0.26479 &
0.28363 &
0.3200 &
2.29 \\
0.8 &
1.30 &
1.800 &
0.23502 &
0.26633 &
0.28416 &
0.3077 &
3.09\\
\end{tabular}
\end{ruledtabular}
\end{table}

\begin{figure}
\includegraphics[angle=0,width=\columnwidth]{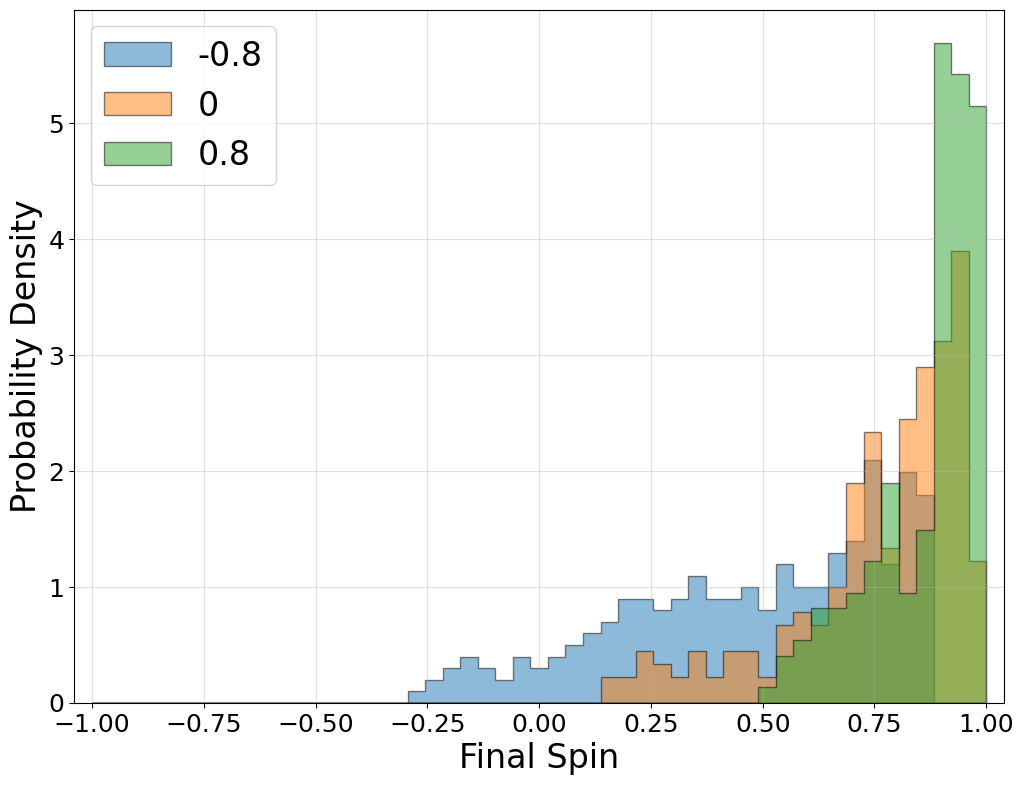}
  \caption{Normalized histograms for the distributions of final spin calculated from the radiated quantities for the three spin families, $s=-0.8, 0.0, +0.8$ using a number of 294, 253, and 222 simulations respectively.
  \label{fig:afhist}}
\end{figure}

While our 769 simulations have been specifically targeted to find the maximum radiated energy covering a large range of black hole spin values,
we can also use this set to find out each remnant black hole spin for those cases.
In Fig.~\ref{fig:afhist} we show in a histogram form the final merged black hole spin $\alpha_f$ as measured
from the radiated quantities for the spin families studied here.
By evaluation of the radiated angular momentum and subtracting from
the initial ADM angular momentum, we can estimate the spin of the remnant black hole.
We clearly see that those binaries with spins aligned with the orbital 
angular momentum lead to the larger remnant spins, as expected.

From the full pull of simulations we extract the one with the largest remnant spin
$\alpha_f=0.9865$ for case $s=+0.80$, $b=1.60$, and $\gamma v=1.15$. In order to
verify this near extremely spinning remnant case 
we perform three additional simulations with increasing resolution as
displayed in Table~\ref{tab:af} showing excellent convergence behavior
and where n$\infty$ is obtained through extrapolation of the n144,
n172, n208 resolutions.

\begin{table}
  \caption{Final remnant black hole horizon mass $m_f^H$ and spin $\alpha_f^H$ of the collision case $s=+0.80$, $b=1.60$, and $\gamma v=1.15$.
    For consistent with the radiative quantities also given
    $M_{ADM}-E_{rad}$ and  $J_{ADM}-J_{rad}$, and convergence order.
  }\label{tab:af} 
\begin{ruledtabular}
\begin{tabular}{ccccc}
Resolution & $m_f^H$ & $\alpha_f^H$ & $\frac{(M_{ADM}-E_{rad})}{m_f^H}$ &  $\frac{(J_{ADM}-J_{rad})}{(m_f^H)^2}$ \\
\hline
n144      & 0.8825 & 0.98599 & 0.8899 & 0.99247\\
n172      & 0.8823 & 0.98636 & 0.8880 & 0.99044\\
n208      & 0.8822 & 0.98650 & 0.8866 & 0.98720\\
n$\infty$ & 0.8821 & 0.98657 & 0.8836 & -- \\
order     & 4.2    & 5.7     & 2.0    & -- \\
\end{tabular}
\end{ruledtabular}
\end{table}

It is worth noting here that unlike the maximum gravitational energy radiated
we search for in this paper, the value of the remnant spin $\alpha_f=0.98657$ 
only represents a lower bound to the maximum spin achievable by these high
energy collisions since we have not seek to fully optimize the parameters
to estimate its maximum. It would be interesting to perform further studies
that include near maximal individual spin holes in the direction of the
orbital angular momentum to challenge the cosmic censorship hypothesis.

We have also look at the maximum final merger black hole spin when we
collide nonspinning holes and have found among our simulations
$\alpha_f^{max}(s=0)\approx0.90$ for $b/M=1.65$ and $\gamma v=1.025$.

\section{Initial scattering and subsequent capture and merger}\label{sec:zw}

In addition to the direct merging cases we have just discussed, in
order to bracket those critical merger impact parameters we performed
several simulations showing an initial scattering into elliptic orbits
(if the overshooting was too large they would continue in hyperbolic orbits).
While in most of our simulations the scattering cases would bounce back
from distances at or below $\sim10M$,
it is of interest to discuss in more detail here a couple of cases bouncing
at larger separations before merging to
illustrate how these can provide additional gravitational radiation besides that
produced at the initial close approach passage.

\begin{figure}
  \includegraphics[angle=0,width=0.7\columnwidth]{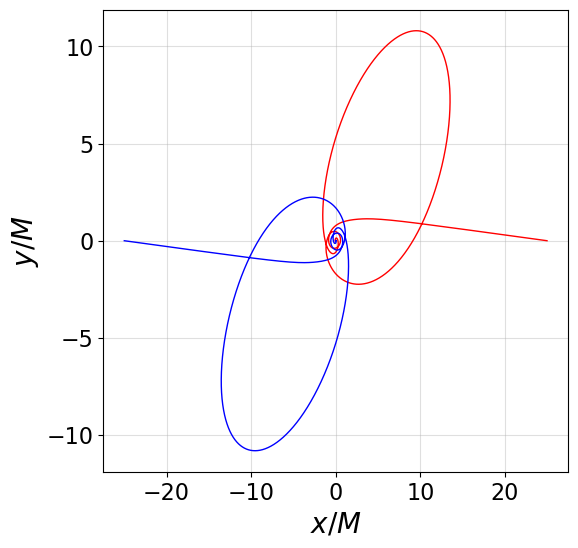}
  \includegraphics[angle=0,width=\columnwidth]{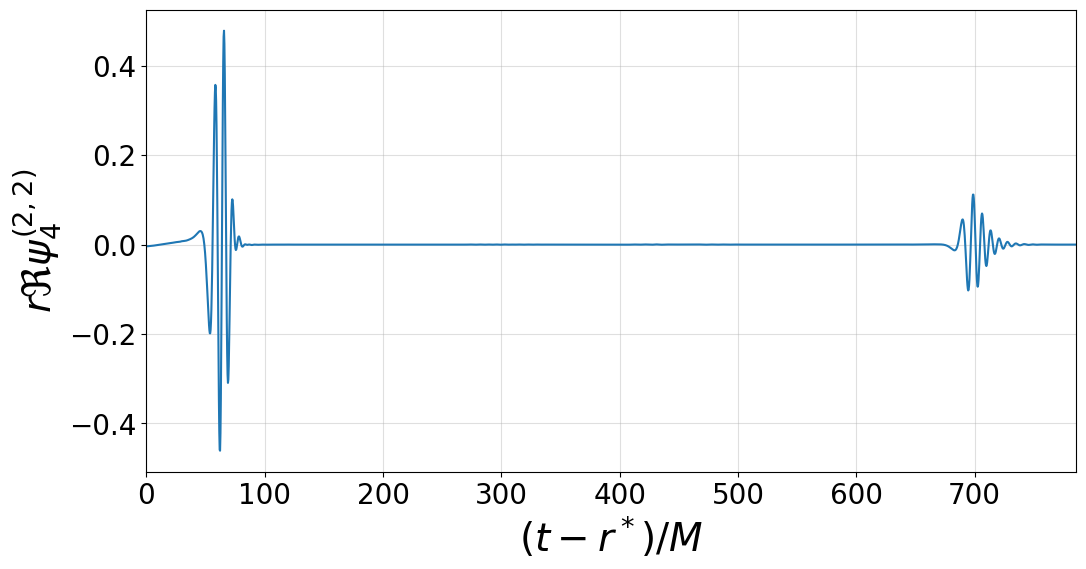}
  \caption{Trajectories in the orbital plane and $\Re\psi_4^{(2,2)}$ (at $R_{obs}=75M)$ for nonspinning binaries with initial momenta parameters $\gamma v=1.45$ and impact parameters $b=1.5$.
  \label{fig:b15gv29}}
\end{figure}

In Fig.~\ref{fig:b15gv29} we display the scattering case with initial momenta parameters $\gamma v=1.45$ and impact parameters $b=1.5$. This configuration approaches until a radius of approximately $0.8M$ before zooming out to about $29M$ and then ultimately merging.
This case has unnormalized $M_{ADM}=1.008$ and individual masses of 0.285, so around 43\% binding energy.
It radiates about 21.6\% in the first pass and 1.5\% afterwards, totaling 23.1\%.

\begin{figure}
  \includegraphics[angle=0,width=0.7\columnwidth]{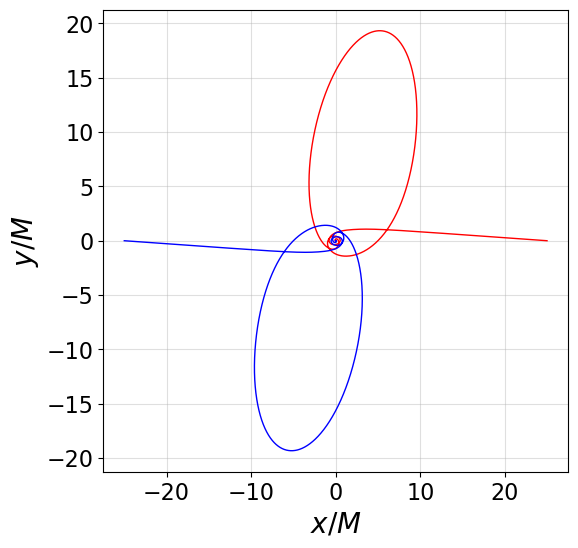}
  \includegraphics[angle=0,width=\columnwidth]{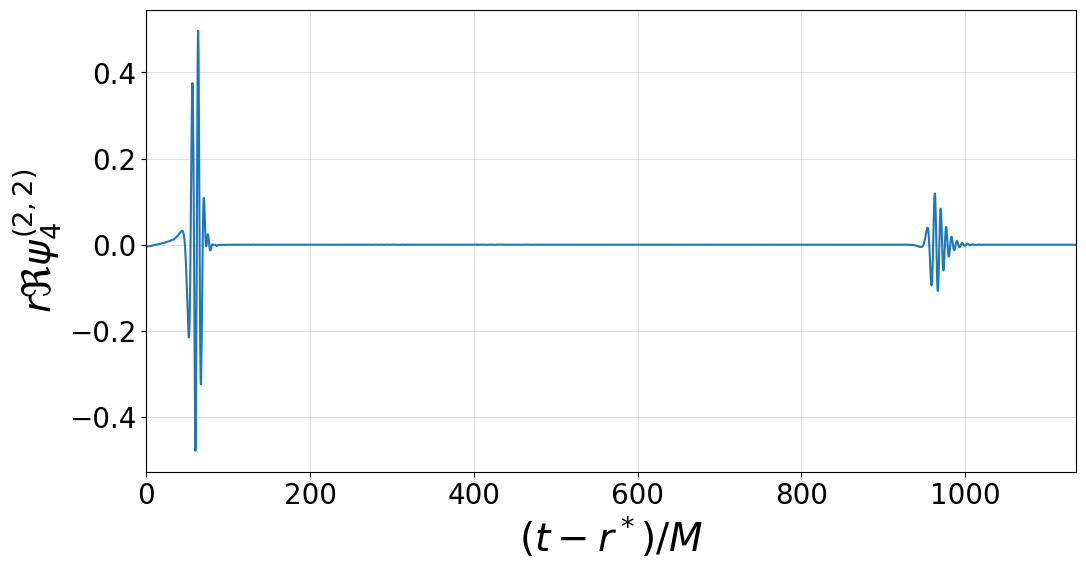}
  \caption{Trajectories in the orbital plane and $\Re\psi_4^{(2,2)}$  (at $R_{obs}=75M)$ for nonspinning binaries with initial momenta parameters $\gamma v=1.76$ and impact parameters $b=1.4$.
  \label{fig:b14gv352}}
\end{figure}

In Fig.~\ref{fig:b14gv352} we display a second scattering case with initial momenta parameters $\gamma v=1.76$, impact parameters $b=1.4$, and $M_{ADM}=1.018$. 
Individual black hole masses are 0.248M each, with the remaining 51\% as binding energy.
In this case, the black holes reach a separation of $0.7M$,
scatter to $41M$, then plunge to merger. This case radiates a total of 23.9\%:  22.5\% in the first approach and 1.4\% in the subsequent merger.

We thus observe that the subsequent radiation by the large eccentricity binary generated by the first encounter
while significant, does not compensate enough to overshoot the values of the optimal direct merger cases.
This might be due to the less efficient radiation compared to the quasicircular orbits at the same binding energy
\cite{Ficarra:2024nro}.

\section{Zero Frequency Limit Expansions}\label{sec:ZFL}

The zero frequency limit (ZFL) approximation was introduced
by Smarr \cite{Smarr:1977fy} for the gravitational radiation emitted during
the high-energy scattering or collision of two black holes.
It was then found that the ZFL not only gives the exact low-frequency results,
but that it provides an estimate of the total energy radiated, its spectrum,
and its angular distribution when applied to the high-velocity collision of
of two particles in a linearized approximation \cite{Payne:1983rrr}.

In this section we expand the computation performed in Ref.~\cite{Berti:2010ce} to higher orders in the impact parameter $b$.
 In our model the two black holes will be treated as point particles initially moving towards each other from the $x$-axes with impact parameter $b$. In the center of momentum frame we can write the coordinates explicitly as

\begin{align}\label{xminf0}
\begin{split}
x_k &= (t,v t \lambda_k,\lambda_k\xi_k,0),\\
p_k &= \gamma_k M_k (1,v \lambda_k,0,0),\\
k &= 1,2 \ \ \ \ \lambda_{1} = 1 \ \ \ \lambda_2 = -1.
\end{split}
\end{align}

with 

\begin{align}\label{xminf01}
\begin{split}
&\xi_1 + \xi_2 = b \\
&M_1 v_1 \gamma_1 = M_2 v_2 \gamma_2
\end{split}
\end{align}

At the closest approach of the holes we set time $t=0$ and the two particles start orbiting around each other in a circular orbit so that the position and momentum of the particles are written as
\begin{align}\label{x0pinf}
\begin{split}
x'_k &= (t,\lambda_k \xi_k \sin{\Omega t},\lambda_k\xi_k \cos{\Omega t},0),\\
p'_k &= (t,\lambda_k v_k \cos{\Omega t},-\lambda_k v_k \sin{\Omega t},0),\\
k &= 1,2 \ \ \ \ \lambda_{1} = 1 \ \ \ \lambda_2 = -1.
\end{split}
\end{align}
with 
\begin{equation}
\Omega = \frac{\gamma_1 M_1 v_1 \xi_1+ \gamma_2 M_2 v_2 \xi_2}{\gamma_1 M_1 \xi_1^2 + \gamma_2 M_2 \xi_2^2}.
\end{equation}

The Fourier transform of the energy-momentum tensor $T^{\mu\nu}(\textbf{k},\omega)$ is written as
\begin{equation}\label{TFourier}
    T^{\mu\nu}(\textbf{k},\omega) = \frac{1}{2\pi} \int d^4x \ T^{\mu\nu}(\textbf{x},t) \  e^{i\left(\omega t - \textbf{k}\cdot\textbf{x}\right)},
\end{equation}
with $T^{\mu\nu}(\textbf{x},t)$ for a point particle given by 
\begin{equation}
T^{\mu\nu}(\textbf{x},t) = \frac{p^\mu_k p^\nu_k}{\gamma_k M_k} \delta(\textbf{x}-\textbf{x}_k(t)).
\end{equation}

Thus we have
\begin{align}\label{energymomentumnopsin}
    \begin{split}
  T^{\mu\nu}(\textbf{k},\omega) &= \int d^4x \frac{p^\mu_k p^\nu_k}{\gamma_k M_k} e^{i(\omega t - \textbf{k}\cdot\textbf{x})}\delta(\textbf{x}-\textbf{x}_k(t))\Theta(-t) \\
  &+ \int d^4x \frac{p'^\mu_k(t) p'^\nu_k(t)}{\gamma_k M_k} e^{i(\omega t - \textbf{k}\cdot\textbf{x})}\delta(\textbf{x}-\textbf{x}'_k(t))\Theta(t)\\ 
  &= \int_{-\infty}^0 dt  \frac{p^\mu_k p^\nu_k}{\gamma_k M_k} e^{i(\omega t - \textbf{k}\cdot\textbf{x}_k(t))} \\
  &+ \int_{0}^\infty dt  \frac{p'^\mu_k(t) p'^\nu_k(t)}{\gamma_k M_k} e^{i(\omega t - \textbf{k}\cdot\textbf{x}'_k(t))}.
    \end{split}
\end{align}

As we have seen from the full numerical simulations, the effect of the spins
of the holes in determining the maximum radiated energy is weak, so for the
sake of the simplicity we will focus here in the nonspinning case.

In this case we have
\begin{align}\label{Erad}
    \begin{split}
        \frac{dE_{rad}}{d\omega d\Omega} = 2\omega^2\left(T^{\mu\nu}T_{\mu\nu}^* - \frac{1}{2}T^\lambda_\lambda T^{\lambda*}_\lambda  \right) .
   \end{split}
\end{align}

While for the linear momentum radiated we have
\begin{equation}\label{Prad}
    \frac{dP_{rad}^i}{d\omega} = \int d\Omega \frac{d^2E_{rad}}{d\omega d\Omega}n_i ,
\end{equation}
where $n_i$ is the unit vector pointing in the i direction.

A similar expression to Eq.~(\ref{Erad}) can be found for the angular momentum which reads
\begin{align}\label{Jrad}
    \begin{split}
        \frac{dJ_{rad}}{d\omega d\Omega} = 2 i \omega\frac{d}{d\phi}\left(T^{\mu\nu}T_{\mu\nu}^* - \frac{1}{2}T^\lambda_\lambda T^{\lambda*}_\lambda  \right).
    \end{split}
\end{align}

We note that it is trivial to show from this expression that the imaginary terms once integrated over the solid angle contribute only as surface terms which are all 0 because the functions involved are periodic on the sphere. 

\subsection{Equal masses}

From the detailed computations described in appendix~\ref{app:zfl}
for the equal mass case $M_1 = M_2 = M/(2\gamma)$, $\xi_1 = \xi_2 = b/2$, $v_1 = v_2 = v$, $\gamma_1 = \gamma_2 = \gamma = (1-v^2)^{-1/2}$, $\Omega = v/\xi$,
the final result for the radiated energy after integrating over the solid angle $\Omega$ is
\begin{widetext}
\begin{align}\label{Enospinequal}
    \begin{split}
    \frac{dE_{rad}}{d\omega}&=\frac{(b\omega) ^6 M^2}{8670412800 \pi  v^7} \bigg[-258048 v^{11}+1507556 v^9-6216210 v^7+13531595 v^5+5358500 v^3\\&-525 \left(v^2-1\right)^2 \left(1396 v^6-3950 v^4+4245
   v^2-3577\right) \tanh ^{-1}(v)-1877925 v\bigg]\\&+\frac{(b\omega)^4 M^2}{2764800 \pi  v^5} \bigg[2560 v^9-10716 v^7+34185 v^5+27270 v^3+45
   \left(v^2-1\right)^2 (204 v^4-281 v^2+195) \tanh ^{-1}{v}-8775 v\bigg]\\&+\frac{(b\omega) ^2 M^2 \bigg[4 v^5+27
   v^3-3 \left(v^2-1\right)^2 \left(4 v^2-3\right) \tanh ^{-1}(v)-9 v\bigg]}{144 \pi  v^3}\\&+\frac{M^2 \bigg[\left(v^4+2 v^2-3\right) \tanh
   ^{-1}{v}-v \left(v^2-3\right)\bigg]}{2 \pi  v}+ \mathcal{O}{(b\omega)^7},
    \end{split}
\end{align}
\end{widetext}
representing an expansion valid for low impact parameter and frequency
combination of the form $b\omega$.

Note that the $v$-dependence can be converted to the $\gamma v$ variable used in the full
numerical simulations by the substitution
$v=(\gamma v)/\sqrt{1+(\gamma v)^2}$.

\section{Fitting ZFL to Simulations}\label{sec:fits}

Here we will try to fit the energy radiated by high energy collisions
with the ZFL dependence on the impact parameter $b$ with coefficients
depending on the linear momentum per mass $\gamma v$ from Eq.~(\ref{Enospinequal}),
which we can thus represent symbolically as
\beq\label{eq:fit}
\frac{E_{rad}}{M_{ADM}}=A f_0(\gamma v)+ B b^2  f_2(\gamma v)+ C b^4
f_4(\gamma v)+ D b^6  f_6(\gamma v), 
\eeq
where the functional form can be read off Eq.~(\ref{Enospinequal}) with
$v=\gamma v/\sqrt{1+(\gamma v)^2}$, and the fitting constants include
an integration over frequencies from $\omega=0$ to an effective one
$\omega_{cuttoff}$.

Notably, when we attempted to fit this dependence to the {229 n100}
simulations
of equal-mass nonspinning high energy collisions, we first observed the
fitting curves produced did not reach above 20\% of radiated energies,
while we have plenty of simulations that reached up to 25\%. We then
reduced successively the simulations to include up to $E_{rad}/M_{ADM}\leq17\%$ and then to
$E_{rad}/M_{ADM}\leq10\%$ not finding statistically very satisfactory fittings until we reduced
to 55 simulations producing $E_{rad}/M_{ADM}\leq5\%$. The results of the fit are
reported in Table~\ref{tab:4parameters}.
These results suggests that the form
of the complete formula applies to small impact parameters only, since
we Taylor expanded the approximation in this $b$ parameter.

\begin{table}
\caption{Four Parameter fit with $E_{rad}/M_{ADM} < 0.05$.  This
fit has 51 degrees of freedom, an rms of 0.00196 and an reduced
chisquare of $3.83e-6$.
\label{tab:4parameters}}
\begin{ruledtabular}
\begin{tabular}{llll}
parameters & fit & Standard & Error\\
\hline
A       &       0.492287     &    +/- 0.02177  &    (4.422\%)\\
B       &       0.438939     &    +/- 0.02512  &    (5.722\%)\\
C       &      -0.0360698    &   +/- 0.009223  &   (25.57\%)\\
D       &       0.00115327   &    +/- 0.000455 &    (39.45\%)\\
\end{tabular}
\end{ruledtabular}
\end{table}

\begin{figure}
  \includegraphics[angle=0,width=\columnwidth]{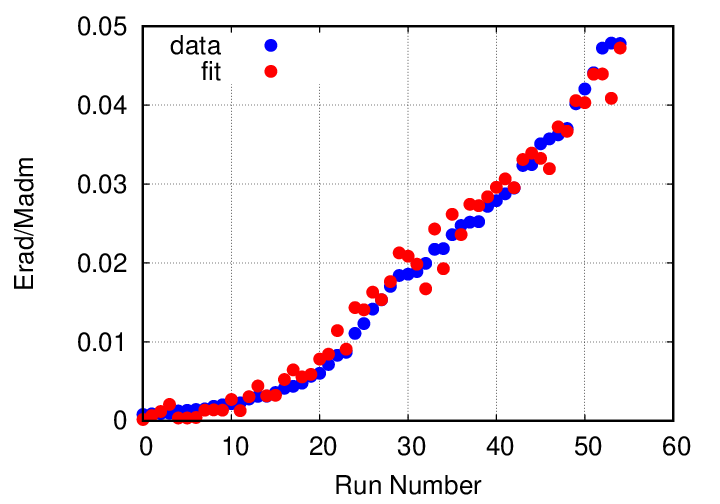}
  \includegraphics[angle=0,width=\columnwidth]{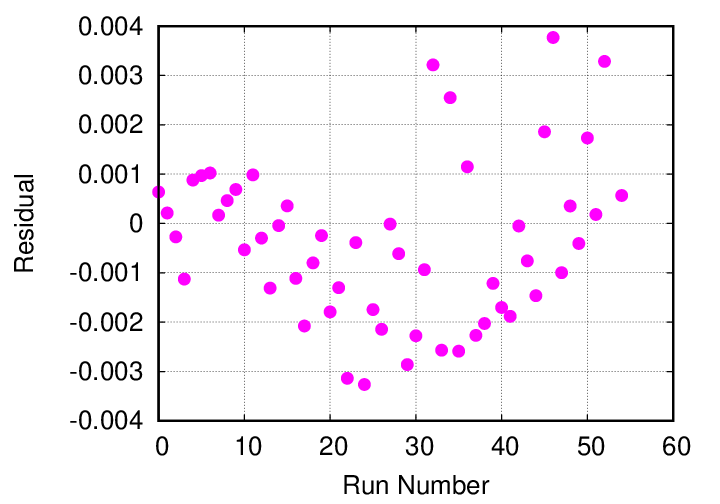}
  \caption{The 4 parameter fit on the restricted data set ($E_{rad}/M_{ADM}<0.05$) compared
    to the actual $E_{rad}$ measured in the 55 simulations for nonspinning high energy
    collisions and its residuals.
  \label{fig:EradLT05}}
\end{figure}

The result of the fitting is also displayed in Fig.~\ref{fig:EradLT05}
were the residuals appear to be within acceptable bounds.

Since we noticed that the fitting value of the $A$ coefficient is very close to
the value 0.5 while that of the coefficient $D$ is very close to 0, we have attempted a new fit by
fixing those values and the result is reported in Table~\ref{tab:2parameters}.
This approach produces a further improvement for the relative errors of fitting parameters
$B$ and $C$.

\begin{table}
\caption{Two Parameter fit with $E_{rad}/M_{ADM} < 0.05$. This 
fit has 53 degrees of freedom, an rms of 0.00205, and a reduced
chisquare of $4.2e-6$.
\label{tab:2parameters}}
\begin{ruledtabular}
\begin{tabular}{llll}
parameters & fit & Standard & Error\\
\hline
B       & 0.417117  &       +/- 0.01174  &    (2.814\%)\\
C       & -0.0200655 &      +/- 0.005073  &   (25.28\%)\\
\end{tabular}
\end{ruledtabular}
\end{table}

A further simplification can be assumed for the final formula by choosing
the simple expressions $A=1/2$, $B=\sqrt{2}-1$, $C=0$, and $D=0$.
Thus producing our final formula as,
\beq\label{eq:fit2}
\frac{E_{rad}}{M_{ADM}}=\frac12 f_0(\gamma v)+ (\sqrt{2}-1) b^2  f_2(\gamma v)\,,
\eeq
where
\bea\label{eq:f0}
f_0(\gamma v)&=&\frac{1}{2\pi}\left[
  \frac{3+2(\gamma v)^2}{1+(\gamma v)^2}+\right.\nonumber\\
  &&-\left.\frac{(3+4(\gamma v)^2)}{(1+(\gamma v)^2)^{3/2}}
  \frac{\ln(\gamma v+\sqrt{1+(\gamma v)^2})}{\gamma v}
  \right]\,,
\eea
and
\bea\label{eq:f2}
f_2(\gamma v)=\frac{1}{144\pi(\gamma v)^2}\left[
  \frac{22(\gamma v)^4+9(\gamma v)^2-9}{1+(\gamma v)^2}+\right.\nonumber\\
   -\left.\frac{(3(\gamma v)^2-9)}{(1+(\gamma v)^2)^{3/2}}
 \frac{\ln(\gamma v+\sqrt{1+(\gamma v)^2})}{\gamma v}
  \right]\,,
\eea
to be applied in the small $(b/M)$ regime where the expansion has been
fit.

\section{Conclusions and Discussion}\label{sec:Discussion}

In this paper we have made a comprehensive search for the maximum gravitational
energy of black hole collisions at relativistic speeds finding an estimate of
of about 1/3 of its total mass.
This result is in rough agreement with
the previous numerical estimates \cite{Sperhake:2009jz} of around 35\%, and
we have also confirmed the
relatively weak dependence on the spins of the black holes of the radiated energy,
in contrast with the strong spin dependence of the net radiated linear momentum \cite{Healy:2022jbh},
leading to a maximum of $v/c\sim1/10$ recoil velocities or the maximum spin of the remnant black hole
as shown in Fig.~\ref{fig:afhist} for aligned spins merging holes.
These results for the maximum possible energy radiated are still below the
45\% of using the area bounds \cite{Eardley:2002re},
final merged black hole horizon mass estimates may be closer to this value
and perhaps even further \cite{Page:2022bem,Page:2022oen} for ultrarelativistic cases.

We have found here that the maximum energy radiated from
relativistic speed collisions lead to (at least) one third of its
total mass, that is a factor about three larger than the maximum
radiated energy from quasicircular orbits
\cite{Scheel:2014ina,Zlochower:2017bbg} for (near) maximally spinning
black holes.  We also recall here that while the high energy collision of
spinning holes lead to a maximum recoil velocity of its merger remnant
of about one tenth the speed of light \cite{Healy:2022jbh}, their
quasicircular merger is limited by about a sixth of this value
\cite{Lousto:2011kp}, i.e. $5000km/s$.

We have been also able to identify in the pool of our initial {739} simulations the one
that lead to the highest final remnant spin and produced a further set of higher
resolution runs with the same $b$ and $\gamma v$ to extract an accurate horizon
measure of $\alpha_f=0.98657$. As our families of simulations did not necessarily
seek for the maximum remnant spin, this represents a lower bound to it. In particular,
this simulation selected one with the spins aligned to the orbital angular
momentum, but with $s=+0.8$ that is still far from the extreme case $s\to1$. This $s\to1$
extreme case cannot be explored with Bowen-York-like conformally flat data,
that is limited to $v<0.9c$ \cite{Healy:2015mla} and 
to $s<0.93$ \cite{Dain:2002ee}, but needs to be explored with the initial data proposed in
\cite{Ruchlin:2014zva,Zlochower:2017bbg} that can reach both, black holes with very high spins
and very highly relativistic speeds and thus may even explore the cosmic censorship hypothesis.
These studies go beyond the scope of the current paper and will be postponed
to a new forthcoming research work.



\begin{acknowledgments}

The authors gratefully acknowledge the National Science Foundation
(NSF) for financial support from Grant No.\ PHY-2207920.
Computational resources were provided by the Blue
Sky, Green Prairies, and White Lagoon clusters at the CCRG-Rochester
Institute of Technology, which were supported by NSF grants
No.\ PHY-1229173, No.\ PHY-1726215, and No.\ PHY-2018420.
This work also used the ACCESS computational resources from the 
allocation PHY060027,
and project PHY20007 in Frontera, an NSF-funded
Petascale computing system at the Texas Advanced Computing Center.
We thanks Robert Caldwell for pointing out typos en Section V.
\end{acknowledgments}


\bibliographystyle{apsrev4-1}
\bibliography{../../../../../Bibtex/references.bib}

\appendix

\section{Zero Frequency limit expressions}\label{app:zfl}

Here we provide explicit expressions for the energy, angular and linear momenta radiated in the ZFL approximation when the two black holes have unequal masses.
Potential applications are to model full numerical simulation of comparable masses and also provide the small mass ratio limit explicitly to compare with other approximations like the self-force approach~\cite{Barack:2018yvs} or the EOB studies~\cite{Nagar:2022icd}.

\subsection{Tensorial computations for the nonspinning case}

In this section we explain in more detail how we evaluated the energy-momentum tensor $T^{\mu\nu}(\textbf{k},\omega)$.

We compute separately the two integrals in Eq.~(\ref{energymomentumnopsin}) as they require different techniques.

Since $p^\mu_k$ is independent from $t$ when $t<0$, the first integral can be computed simply as 
\begin{align}
  \begin{split}
&\int_{-\infty}^0 dt  \frac{p^\mu_k p^\nu_k}{\gamma_k M_k} e^{i(\omega t - \textbf{k}\cdot\textbf{x}_k(t))}=\\
&=\frac{p^\mu_k p^\nu_k}{\gamma_k M_ki\omega(1-v_k \lambda_k \sin{\theta}\cos{\phi})}e^{-i\lambda_k\xi_k\omega\sin{\theta}\sin{\phi}}.
    \end{split}
\end{align}

For $t>0$
\begin{align}\label{secondintegral}
    \begin{split}
  &\int d^4x \frac{p^{'\mu}_k(t) p^{'\nu}_k(t)}{\gamma_k M_k} e^{i(\omega t - \textbf{k}\cdot\textbf{x})}\delta(\textbf{x}-\textbf{x}^{'}_k(t))\Theta(t)= \\&
  = \int^{\infty}_0 dt  \frac{p^{'\mu}_k(t) p^{'\nu}_k(t)}{\gamma_k M_k} e^{i(\omega t - \textbf{k}\cdot\textbf{x}^{'}_k(t))}.
    \end{split}
\end{align}

The exponential in Eq.~(\ref{secondintegral}) can be expanded in Bessel functions as
\begin{align}
\begin{split}
   &e^{i\omega t -i\textbf{k}\cdot \textbf{x}^{'}_k(t)} = e^{i\omega t} \exp\left\{- i\lambda_k \omega \xi_k \sin\theta \sin(\Omega t + \phi)\right\}=\\&
   =  e^{i\omega t} \sum_{n=-\infty}^{n=\infty} J_n(-\lambda_k \omega \xi_k \sin\theta) e^{in\Omega t + in\phi},
\end{split}
\end{align}
so that Eq.~(\ref{secondintegral}) becomes
\begin{align}
    \begin{split}
 &\int^{\infty}_0 dt  \frac{p^{'\mu}_k(t) p^{'\nu}_k(t)}{\gamma_k M_k} e^{i(\omega t - \textbf{k}\cdot\textbf{x}^{'}_k(t))} =
 \\&=\sum_{n=-\infty}^{n=\infty} J_{n}(-\lambda_k \omega \xi_k \sin\theta) \int^{\infty}_0 dt  \frac{p^{'\mu}_k(t) p^{'\nu}_k(t)}{\gamma_k M_k}  e^{i\omega t} e^{in(\Omega t + \phi)} \\
    \end{split}.
\end{align}

To work out the contribution given to $T^{\mu\nu}(\textbf{k},\omega)$ by $T^{\mu\nu}_{tens}(\textbf{x},t)$ in Eq.~(28) of Ref.~\cite{Berti:2010ce} we use the fact that $T^{\mu\nu}_{tens}(\textbf{x},t)$ can be written as
\begin{equation}
    T^{\mu\nu}_{tens}(\textbf{x},t) =   \tilde{T}^{\mu\nu}(t) \delta(\cos{\theta})\Theta(t)\sum_{k=1}^2 \tau_k(r)\delta(\phi + \Omega t - \phi_k),
\end{equation}
where $\tau_k(r)$ is given by Eq.~(27) of Ref.~\cite{Berti:2010ce}, $ \tilde{T}^{\mu\nu}(t)$ is some tensor depending only only on t and $\phi_1 = \frac{\pi}{2}$ and $\phi_2=\frac{3}{2}\pi$.

Now, using that 
\begin{align}
    \begin{split}
  &e^{i\omega t -i\textbf{k}\cdot \textbf{x}'} =\\
  &e^{i\omega t} \exp\left\{- ir\omega \sin{\theta'}\sin{\theta}\cos(\phi - \phi') - ir\omega \cos{\theta}\cos{\theta'}\right\},
    \end{split}
\end{align}
we find
\begin{align}
    \begin{split}
    &\frac{1}{2\pi} \int d^4x' T^{\mu\nu}_{tens}(\textbf{x},t)   e^{i\omega t -i\textbf{k}\cdot \textbf{x}}=
    \\&= \frac{1}{2\pi}\sum_{k=1}^2\int_0^\infty dt' \  \tilde{T}^{\mu\nu}(t)\int_0^{\infty} dr' \ r'^2 \tau_k(r') \int_{-1}^1 dx' \delta(x')
    \\&\times \int_0^{^2\pi} d\phi' \delta(\phi' + \Omega t - \phi_k)   e^{i\omega t -i\textbf{k}\cdot \textbf{x}'} \\
    &= \frac{1}{2\pi} \sum_{k=1}^2\int_0^\infty dt \  \tilde{T}^{\mu\nu}(t) e^{i\omega t} 
    \\& \times \int_0^{\infty} dr \ r^2 \tau_k(r)  \exp\left\{- ir\omega \sin{\theta}\cos((\phi + \Omega t) - \phi_k)\right\} \\
    &= \frac{1}{2\pi} \sum_{k=1}^2 \frac{M_k \xi_k \Omega^2}{\sqrt{1-(\Omega\xi_k)^2}}\int_0^\infty dt \  \tilde{T}^{\mu\nu}(t) e^{i\omega t}\\& \times \int_0^{\xi_k} dr \exp\left\{- ir\lambda_k\omega \sin{\theta}\sin(\phi + \Omega t)\right\}.
    \end{split}
\end{align}

Using again the previous expansion in Bessel functions we find
\begin{align}
    \begin{split}
    &\frac{1}{2\pi} \int d^4x' T^{\mu\nu}_{tens}(\textbf{x},t)   e^{i\omega t -i\textbf{k}\cdot \textbf{x}}\\& = \frac{1}{2\pi} \sum_{k=1}^2 \frac{M_k \xi_k \Omega^2}{\sqrt{1-(\Omega\xi_k)^2}} \sum_{n=-\infty}^{\infty}\left(\int_0^\infty dt \  \tilde{T}^{\mu\nu}(t) e^{i\omega t + i n(\Omega t + \phi)}\right)\\& \times \int_0^{\xi_k} dr J_n(-r\lambda_k\omega \sin{\theta}),
    \end{split}
\end{align}
since the components $T^{\mu t}(\textbf{k},\omega)$ can be computed from 
\begin{equation}\label{Ttt}
    T^{t t}(\textbf{k},\omega) = \hat{k}^i\hat{k}^j T_{ij}(\textbf{k},\omega),
\end{equation}
and
\begin{equation}\label{Tti}
    T^{t i}(\textbf{k},\omega) = \hat{k}^j T_{ij}(\textbf{k},\omega).
\end{equation}

Thus we can focus on the computation of only the spatial components of the energy-momentum tensor and our final expression reads
\begin{align}
    \begin{split}
      &T^{ij}(\textbf{k},\omega) =\\
      &\frac{1}{2\pi}\sum_{k=1}^2 \Bigg\{\frac{p^i_k p^j_k}{\gamma_k M_k i\omega(1-v_k\lambda_k \sin{\theta}\cos{\phi})}e^{-i\lambda_k\xi_k\omega\sin{\theta}\sin{\phi}}\\
        &+\sum_{n=-\infty}^{n=\infty} J_{n}(-\lambda_k \omega \xi_k \sin\theta) \int^{\infty}_0 dt  \frac{p^{'i}_k(t) p^{'j}_k(t)}{\gamma_k M_k}  e^{i\omega t} e^{in(\Omega t + \phi)}\\
       & +\frac{M_k \xi_k\Omega^2}{\sqrt{1-(\Omega\xi_k)^2}} \bigg(\sum_{n=-\infty}^{\infty}  \left(\int_0^\infty dt \  \tilde{T}^{ij}(t) e^{i\omega t} e^{in(\Omega t + \phi)}\right)  \\& \times\int_0^{\xi_k} dr J_n(-r\lambda_k\omega \sin{\theta})\Bigg\},
    \end{split}
\end{align}
with
\begin{align}
    \begin{split}
            &\tilde{T}^{xx}(t)= -\sin^2(\Omega t),\\
            &\tilde{T}^{yy}(t)= -\cos^2(\Omega t),\\
            &\tilde{T}^{xy}(t)= -\sin(\Omega t)\cos(\Omega t).
    \end{split}
\end{align}

This can then be used in Eqs.~(\ref{Erad})-(\ref{Jrad}) to compute radiated quantities as follows next.

\subsection{General case}

For generic masses and velocities we find once integrating over the solid angle
Eqs.~(\ref{Erad})-(\ref{Jrad})
\begin{widetext}
\begin{align}
    \begin{split}
     \frac{dE_{rad}}{d\omega} &=  \frac{(b\omega) ^4 (\gamma_1 M_1 v_1)^2}{345600 \pi  v_1^6 v_2^6 (v_1+v_2)^3} \Bigg\{2 v_1 v_2 (v_1+v_2) \bigg[2560 v_1^9 v_2^5-4 v_1^8 (1280
   v_2^6+459 v_2^4+270 v_2^2-405)\\
   &+12 v_1^7 v_2 (640 v_2^6+763 v_2^4+1480
   v_2^2-1320)+v_1^6 (-5120 v_2^8-25356 v_2^6-30135 v_2^4+21510 v_2^2+2025)\\&+v_1^5
   v_2 (2560 v_2^8+9156 v_2^6+61095 v_2^4+98620 v_2^2-19050)-v_1^4 v_2^2 (1836v_2^6+30135 v_2^4+184550 v_2^2-25275)\\&+5 v_1^3 v_2^3 (3552 v_2^4+19724
   v_2^2-5055)+15 v_1^2 v_2^4 (-72 v_2^4+1434 v_2^2+1685)-30 v_1 v_2^5 (528
   v_2^2+635)+405 v_2^6 (4 v_2^2+5)\bigg]\\&-30 v_1^6 (v_2^2-1)^2 (108 v_1^4
   (v_2^2+1)-12 v_1^3 v_2 (39 v_2^2+79)+9 v_1^2 (12 v_2^4+67
   v_2^2+15)+v_1 v_2 (-1188 v_2^4+6117 v_2^2-1135)\\&+v_2^2 (828 v_2^4-5037
   v_2^2+415)) \tanh ^{-1}{v_2}-30 (v_1^2-1)^2 v_2^6 (108 (v_1^2+1)
   v_2^4-12 v_1 (39 v_1^2+79) v_2^3\\&+9 (12 v_1^4+67 v_1^2+15) v_2^2+v_1
   (-1188 v_1^4+6117 v_1^2-1135) v_2+v_1^2 (828 v_1^4-5037 v_1^2+415)) \tanh
   ^{-1}(v_1)\Bigg\}\\&+\frac{(b\omega) ^2 (\gamma_1 M_1 v_1)^2}{36 \pi 
   v_1^4 v_2^4 (v_1+v_2)} \bigg[v_1 v_2
   (2 v_1^5 (5 v_2^2-3)-6 v_1^4 v_2 (v_2^2-4)+v_1^3 (-6 v_2^4+9
   v_2^2-9)+v_1^2 v_2^3 (10 v_2^2+9)+24 v_1 v_2^4\\&-3 v_2^3 (2
   v_2^2+3))+3 (2 (v_1^2-4 v_1 v_2+v_2^2)+3) (v_1^4
   (v_2^2-1)^2 \tanh ^{-1}{v_2}+(v_1^2-1)^2 v_2^4 \tanh ^{-1}{v_1})\bigg]\\
   &+\frac{2 (\gamma_1 M_1 v_1)^2}{\pi  v_1^3 v_2^3 (v_1+v_2)} \bigg[-v_1 v_2 (v_1+v_2)
   \Big((v_1^2-1) v_2^2-v_1^2-v_1 v_2\Big)+v_1^3 (v_2^2-1) (v_1
   v_2^2+v_1+2 v_2) \tanh ^{-1}{v_2}\\
   &+(v_1^2-1) v_2^3 (v_1 (v_1
   v_2+2)+v_2) \tanh ^{-1}{v_1}\bigg]+\mathcal{O}\left((b\omega) ^5\right),
    \end{split}
\end{align}
\begin{align}
    \begin{split}
     \frac{dP_{rad}^x}{d\omega}&=\frac{(\gamma_1 M_1 v_1)^2}{\pi  v_1^4 v_2^4 (v_1+v_2)} \bigg[v_1 v_2 (v_1-v_2) (v_1+v_2) \Big(v_1^2 (v_2^2-3)-5
   v_1 v_2-3 v_2^2\Big)\\
   &+v_1^4 (v_2^2-1) (-v_1 (v_2^2+3)+v_2^3-5
   v_2) \tanh ^{-1}{v_2}+(v_1-1) (v_1+1) v_2^4 \Big(v_1 (v_1 (v_2-v_1)+5)+3 v_2\Big)
   \tanh ^{-1}{v_1}\bigg]\\
   &-\frac{(b\omega) ^2 (\gamma_1 M_1 v_1)^2}{36 (\pi  v_1^5 v_2^5 (v_1+v_2))}\bigg[ \big(2
   (v_1^2-4 v_1 v_2+v_2^2)+3\big)\\
   &(v_1^5 v_2 (5 v_2^2-3)-5 v_1^3
   v_2^5+3 v_1^5 (v_2^2-1)^2 \tanh ^{-1}(v_2)-3 (v_1^2-1)^2 v_2^5 \tanh
   ^{-1}{v_1}+3 v_1 v_2^5)\bigg]\\
   &+\frac{(b\omega) ^4
   (\gamma_1 M_1 v_1)^2}{345600 \pi  v_1^7
   v_2^7 (v_1+v_2)^3} \bigg\{2 v_1 v_2 (v_1-v_2) (v_1+v_2) \Big[12 v_1^8 (1280 v_2^6+153
   v_2^4+90 v_2^2-135)\\
   &-12 v_1^7 v_2 (1280 v_2^6+343 v_2^4+1390 v_2^2-1185)+3
   v_1^6 (5120 v_2^8-34188 v_2^6+4485 v_2^4-2430 v_2^2-675)\\
   &+v_1^5 v_2 (-4116
   v_2^6+163145 v_2^4-105910 v_2^2+17025)+v_1^4 v_2^2 (1836 v_2^6+13455 v_2^4+78640
   v_2^2-8250)\\
   &-5 v_1^3 v_2^3 (3336 v_2^4+21182 v_2^2-3405)\\
   &+30 v_1^2 v_2^4 (36
   v_2^4-243 v_2^2-275)+15 v_1 v_2^5 (948 v_2^2+1135)-405 v_2^6 (4
   v_2^2+5)\Big]\\
   &+30 v_1^7 (v_2^2-1)^2 (108 v_1^4 (v_2^2+1)-12 v_1^3
   v_2 (39 v_2^2+79)+9 v_1^2 (12 v_2^4+67 v_2^2+15)\\
   &+v_1 v_2 (-1188
   v_2^4+6117 v_2^2-1135)+v_2^2 (828 v_2^4-5037 v_2^2+415)) \tanh ^{-1}{v_2}\\
   &-30(v_1^2-1)^2 v_2^7 (108 (v_1^2+1) v_2^4-12 v_1 (39 v_1^2+79)
   v_2^3+9 (12 v_1^4+67 v_1^2+15) v_2^2\\
   &+v_1 (-1188 v_1^4+6117 v_1^2-1135)
   v_2+v_1^2 (828 v_1^4-5037 v_1^2+415)) \tanh ^{-1}{v_1}\bigg\}+\mathcal{O}((b\omega) ^5),
    \end{split}
\end{align}
\begin{align}
    \begin{split}
     \frac{dJ^z_{rad}}{d\omega}&=\frac{b (\gamma_1 M_1 v_1)^2}{\pi  v_1^3 v_2^3
   (v_1+v_2)^2}\bigg[-v_1 v_2 (v_1+v_2) \Big(v_1^3 v_2 (v_2^2-3)+v_1^2
   (1-6 v_2^2)+v_1 (v_2-3 v_2^3)+v_2^2\Big)\\&+v_1^3 (v_2^2-1)
   \Big(v_1^2 v_2 (v_2^2+3)+v_1 (5 v_2^2-1)+2 v_2 (v_2^2-1)\Big)
   \tanh ^{-1}{v_2}\\&+(v_1^2-1) v_2^3 (v_1 (v_1^2 (v_2^2+2)+5 v_1
   v_2+3 v_2^2-2)-v_2) \tanh ^{-1}{v_1}\bigg]\\&-\frac{(b\omega) ^2 b (\gamma_1 M_1 v_1)^2}{72 \pi  v_1^4 v_2^4
   (v_1+v_2)^2} \bigg[v_1 v_2 (v_1+v_2) \Big(32 v_1^5
   v_2^3+v_1^4 (-41 v_2^4-24 v_2^2+9)+v_1^3 v_2 (32 v_2^4+27 v_2^2-135)\\&-3
   v_1^2 (8 v_2^4-87 v_2^2+9)+27 v_1 v_2 (1-5 v_2^2)+9 v_2^2
   (v_2^2-3)\Big)+9 v_1^4 (v_2^2-1)^2 \Big(v_1^2 (v_2^2-1)+14 v_1
   v_2-9 v_2^2+3\Big) \tanh ^{-1}{v_2}\\&+9 (v_1^2-1)^2 v_2^4 (v_1^2
   (v_2^2-9)+14 v_1 v_2-v_2^2+3) \tanh ^{-1}{v_1}\bigg]+\mathcal{O}((b\omega) ^5).
    \end{split}
\end{align}

\end{widetext}

Note that these formulas have been found in the initial center of momentum frame
as given in Eq.~(\ref{xminf0}),
and there verify the exchange of the black hole labeling 1 and 2 symmetry.

\end{document}